\documentclass[aps,pra,reprint,superscriptaddress,showkeys,amsmath,amssymb]{revtex4-2}
\usepackage[utf8]{inputenc}
\usepackage{graphicx}
\usepackage{physics}
\usepackage[hidelinks]{hyperref}

\makeatletter
\def\maketitle{
\@author@finish
\title@column\titleblock@produce
\suppressfloats[t]
}
\renewcommand{\fnum@figure}{\textbf{Fig.\,\thefigure}}
\def\@caption@fignum@sep{\textbf{:}~}
\makeatother
\begin{document}


\author{R. Barczyk}
\affiliation{Center for Nanophotonics, AMOLF, Science Park 104, 1098 XG Amsterdam, The Netherlands}
\author{L. Kuipers}
\affiliation{Kavli Institute of Nanoscience, Delft University of Technology, 2600 GA, Delft, The Netherlands}
\author{E. Verhagen}\email{e.verhagen@amolf.nl}
\affiliation{Center for Nanophotonics, AMOLF, Science Park 104, 1098 XG Amsterdam, The Netherlands}

\title{Observation of Landau levels and topological edge states in photonic crystals through pseudomagnetic fields induced by synthetic strain}

\begin{abstract}

The control over light propagation and localization in photonic crystals offers wide applications from sensing and on-chip routing to lasing and quantum light-matter interfaces. While in electronic crystals magnetic fields can be used to induce a multitude of unique phenomena, the uncharged nature of photons necessitates alternative approaches to bring about similar control over photons at the nanoscale. Here, we experimentally realize pseudomagnetic fields in two-dimensional photonic crystals through engineered strain of the lattice. Analogous to strained graphene, this induces flat-band Landau levels at discrete energies. We study the spatial and spectral properties of these states in silicon photonic crystals at telecom wavelengths with far-field spectroscopy. Moreover, taking advantage of the photonic crystal's design freedom, we realize domains of opposite pseudomagnetic field and observe topological edge states at their interface. We reveal that the strain-induced states can achieve remarkably high quality factors despite being phase-matched to the radiation continuum. Together with the high density of states and high degeneracy associated with flat bands, this provides powerful prospects for enhancing light-matter interactions, and demonstrates a design principle to govern both on-chip and radiating light fields.

\end{abstract}

\maketitle

\begin{figure*}[hbt]
  \centering
  \includegraphics[width=\textwidth]{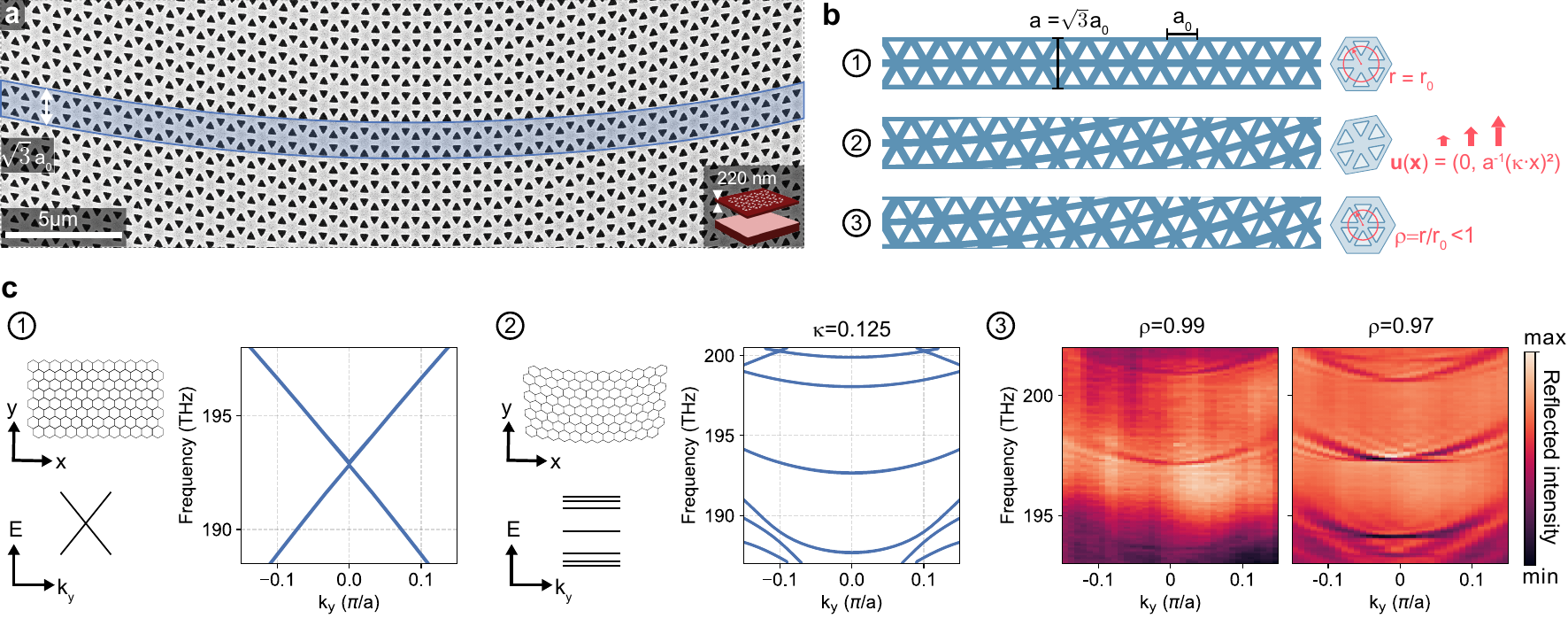}
  \caption{\textbf{Sample geometry and strain-induced Landau levels.} \textbf{a,} Scanning electron micrograph of a fabricated photonic crystal membrane, with the sample geometry depicted in the inset. The supercell of the strained honeycomb lattice is highlighted in blue, with periodicity $a=\sqrt{3}a_0$ along the y-direction, and $a_0$ denoting the lattice constant of the underlying pristine honeycomb lattice. \textbf{b,} Schematic depiction of supercells for the (1) pristine, (2) strained, and (3) shrunken strained lattices. The applied displacement function is given by $\mathbf{u}(\vb{x})$, and $\rho=r/r_0$ denotes the relative displacement of holes within a hexagonal unit cell with respect to a pristine lattice ($\rho=1$). \textbf{c,} 1: Schematic real-space geometry and dispersion of a pristine honeycomb lattice (left), together with simulated photonic bulk bands. The dispersion is characterized by a Dirac-type linear crossing. 2: A bandgap is opened in place of the original Dirac-type crossing due to strain, and flat Landau levels emerge symmetrically distributed around the bandgap center. 3: Experimentally retrieved angularly resolved reflection spectra showing photonic Landau levels in strained photonic crystal lattices with $\rho=0.99$ and $0.97$ ($\kappa=0.125$ in both cases).}
  \label{fig:Fig1}
\end{figure*}

In condensed matter physics, magnetic fields provide a versatile mechanism to control the behavior of electrons in materials. For example, a magnetic field piercing a two-dimensional electron gas induces flat bands at discrete energies known as Landau levels, which can be viewed as the quantization of the electrons' cyclotron motion in the magnetic field. Moreover, at the system's boundaries, the magnetic field implies the existence of topological edge states associated with the quantum Hall effect. In photonics, the idea of controlling light in a similar way has been a tantalizing prospect. While at microwave frequencies magneto-optic effects may be strong enough to effectively mediate interactions between photons and real magnetic fields~\cite{wangObservationUnidirectionalBackscatteringimmune2009}, this approach is unfeasible for optical frequencies. Realizing an effective magnetic control over photons in dielectric photonic systems without relying on actual external magnetic fields represents an especially luring proposition for on-chip, nanophotonic systems.
In graphene, pseudomagnetic fields (PMFs) can be induced for electrons via mechanical strain of the lattice, as the corresponding perturbation to inter-atomic hopping mimics the action of a magnetic gauge potential \cite{kaneSizeShapeLow1997, guineaEnergyGapsZerofield2010,levyStrainInducedPseudoMagnetic2010, gomesDesignerDiracFermions2012}. Contrary to real magnetic fields that break time-reversal symmetry, strain-induced PMFs carry opposite signs for the two non-equivalent Dirac cones at the K and K' valleys. Nevertheless, they still give rise to intriguing phenomena including Landau-level quantization and topological edge states.
Analogously, suitable lattice deformations can act as magnetic gauge potentials in bosonic systems. Their effects have been studied in lattices of coupled waveguides \cite{rechtsmanStraininducedPseudomagneticField2013, songDispersionlessCouplingOptical2022}, arrays of microwave resonators \cite{bellecObservationSupersymmetricPseudoLandau2020, jiaExperimentalRealizationChiral2022}, microcavity exciton-polaritons \cite{jamadiDirectObservationPhotonic2020, lledoPolaritonCondensationVortex2022}, and in acoustic platforms~\cite{zhengLandauLevelsVan2021, wenAcousticLandauQuantization2019, yangStrainInducedGaugeField2017, abbaszadehSonicLandauLevels2017, periAxialfieldinducedChiralChannels2019}. Recently, Guglielmon and coworkers predicted that inhomogeneous deformations in honeycomb photonic crystal (PhC) membranes can act as a magnetic gauge potential, creating synthetic strain through a designed perturbation of the dielectric function $\epsilon(\mathbf{x})$~\cite{guglielmonLandauLevelsStrained2021}.
Strain-induced pseudomagnetism in photonic crystals is nontrivial because of inherent long-range interactions. It would however be highly appealing as it provides a new paradigm for the on-chip routing and confinement of light~\cite{salernoHowDirectlyObserve2015, salernoPropagatingEdgeStates2017}, and a path to enhance light-matter interactions and nonlinearities through the high degeneracy and high local density of states of photonic Landau levels, associated with their nature as flat bands~\cite{borregaardQuantumNetworksDeterministic2019,kraussWhyWeNeed2008, smirnovaNonlinearTopologicalPhotonics2020, schomerusParityAnomalyLandauLevel2013, yangPhotonicFlatbandResonances2023}.
Here, we realize pseudomagnetic fields in silicon photonic crystals at telecom wavelengths, and demonstrate the emergence of photonic Landau levels. Using far-field Fourier spectropolarimetry~\cite{gorlachFarfieldProbingLeaky2018, parappurathDirectObservationTopological2020, barczykInterplayLeakageRadiation2022}, we study the characteristic energy scaling of Landau levels with strain, their delocalization, and loss mechanisms. We reveal that they can exhibit remarkably high quality factors, even though they exist within the radiation continuum. We finally demonstrate the creation of inhomogeneous pseudomagnetic fields by spatial tailoring of the strain field. This allows observing signatures of the topological edge states that are predicted to exist at the boundaries of domains with opposite magnetic fields~\cite{huangPatterntunableSyntheticGauge2022} and which are distinct from other crystal symmetry-based implementations of topological edge states in photonic crystals~\cite{wuSchemeAchievingTopological2015, salernoPropagatingEdgeStates2017, kiriushechkinaSpindependentPropertiesOptical2023, renZeroenergyEdgeStates2023}. Our findings illustrate the applicability of synthetic strain engineering for the control of light and its interaction with matter at the nanoscale.

\section*{Results}

\subsection*{Observation of photonic Landau levels}
We fabricate suspended silicon membranes perforated by triangular air holes using electron-beam lithography and wet etching of a silicon-on-insulator substrate (see Methods)~\cite{barikTwodimensionallyConfinedTopological2016, reardonFabricationCharacterizationPhotonic2012}. Figure~\ref{fig:Fig1}a shows a scanning electron micrograph of a fabricated strained PhC lattice, where the inset depicts a three-dimensional cross-cut of the slab. We follow the approach outlined in  \cite{guglielmonLandauLevelsStrained2021} to induce a uniform pseudomagnetic field $\vb{B}_\text{eff}$ piercing the PhC plane. The starting point is a pristine TE-type honeycomb PhC with lattice constant $a_0$ and dielectric distribution $\epsilon (\vb{x})$ ($\vb{x}=(x,y)$). The lattice is oriented with the zigzag and armchair directions along the $x$ and $y$ axes, respectively. The PhC's frequency spectrum features a Dirac-type crossing (Fig.~\ref{fig:Fig1}b,c) and is governed by the scalar Helmholtz equation for the out-of-plane magnetic field $\vb{H}(\vb{x}) = H_z(\vb{x}) \vb{\hat{z}}$
\begin{equation}
\label{eq:helmholtz}
- \grad \cdot (\epsilon^{-1}(\vb{x})\ \grad) H_z(\vb{x}) = (\omega/c)^2\ H_z(\vb{x}) \ .
\end{equation}
The calculated dispersion is plotted in Fig.~\ref{fig:Fig1}c for the $k_y$ direction, with the Dirac point appearing at $k_y=0$ for lattice constant $a=\sqrt{3}a_0$, which is the minimal periodicity when implementing the strain profile.
\begin{figure*}[hbt]
  \centering
  \includegraphics[width=\textwidth]{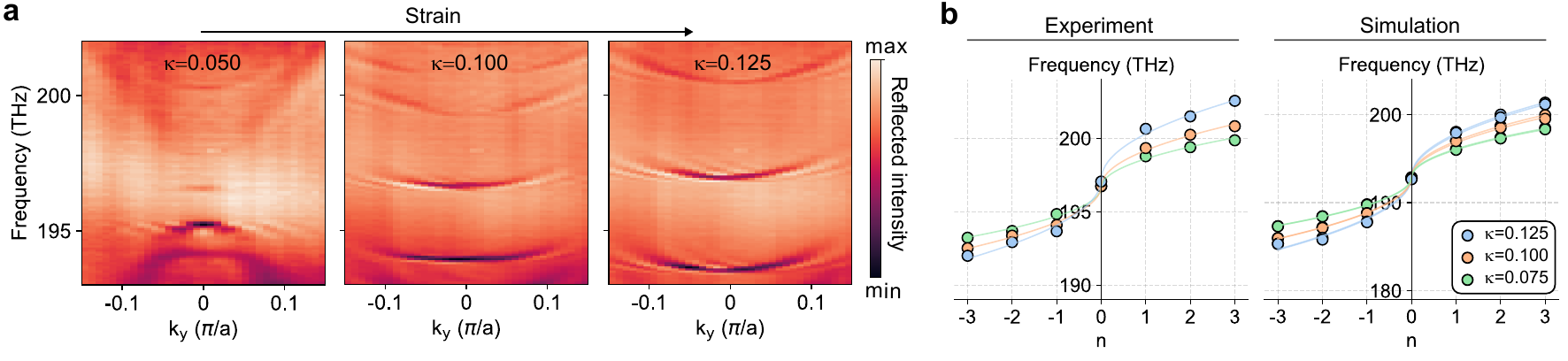}
  \caption{\textbf{Photonic Landau levels in increasingly strong pseudomagnetic fields.} \textbf{a,} Far-field reflection spectra displaying the evolution of photonic bands with increasing strain magnitude $\kappa$. \textbf{b,} Experimentally and numerically extracted center mode frequencies of photonic Landau levels for varying $\kappa$, fitted respectively with a square-root scaling law. The small mode splitting of the (ideally doubly degenerate) modes in simulations is a finite-size effect and not resolvable in our far-field measurements.}
  \label{fig:Fig2}
\end{figure*}
This uniaxial strain is realized by displacing each point $\vb{x}$ in the PhC plane via $T(\vb{x}) = \vb{x}+\vb{u}(\vb{x})$, with quadratic displacement function $\vb{u}(\vb{x}) = (0, a^{-1}(\kappa x)^2)$ and the parameter $\kappa$ representing the strain magnitude. The displacement thus breaks the $x$-periodicity but preserves the larger super-cell periodicity $a$ along $y$.
In the vicinity of the K and K' valleys the eigenfunctions' dynamics are captured by a 2D Dirac-type Hamiltonian \cite{guineaEnergyGapsZerofield2010}
\begin{equation}
\label{eq:hamilton}
H_\text{eff} = v_D[-i\grad_{\vb{x}} - \vb{A}_{\text{eff}}(\vb{x})]\cdot\vb{\sigma}
\end{equation}
with Dirac velocity $v_D$, Pauli matrices $\vb{\sigma}=(\sigma_1, \sigma_2)$, and effective magnetic vector potential $\vb{A}_{\text{eff}}$ that is related to the displacement function $\vb{u}$ via
\begin{equation}
  \label{eq:aeff}
  \vb{A}_{\text{eff}} \propto 
  \begin{pmatrix}
           u_{x,x}-u_{y,y} \\
           -(u_{x,y}+u_{y,x})
   \end{pmatrix} \ ,
\end{equation}
where $u_{i,j}=\frac{\partial u_i}{\partial x_j}$. The solutions to the eigenvalue problem associated with equation~(\ref{eq:hamilton}), up to first order $\mathcal{O}(\kappa)$, correspond to flat states at discrete frequencies that are symmetrically distributed around the original Dirac-cone frequency $\omega_D$. These constitute Landau levels that follow the square-root law
\begin{equation}
\label{eq:landau}
\omega_n = \omega_D \pm \frac{v_D c^2}{\sqrt{2}\omega_D} \sqrt{n |\vb{B}_{\text{eff}}(\kappa)|}, \quad n=0,1,2,... .
\end{equation}
where the effective magnetic field amplitude is given by $|\vb{B}_{\text{eff}}(\kappa)|= |\grad_{\vb{x}}\cross \vb{A}_{eff}(\kappa)|=B_0 \kappa^2$, and $B_0$ is a constant parameter specific to the (meta)material~\cite{guglielmonLandauLevelsStrained2021, huangPatterntunableSyntheticGauge2022}. 

To make the strain-induced photonic Landau levels experimentally accessible via far-field Fourier spectroscopy (see Supplemental Material Fig.~\ref{fig:FigED1} and Methods for details on the setup), we enhance their radiative coupling by applying an additional type of sub-lattice symmetry breaking via concentrically shrinking the radial position $r$ of six air holes within a hexagonal unit cell by a factor of $\rho = r/r_0$ (Fig.~\ref{fig:Fig1}b,c), where $r=r_0$ for the pristine honeycomb PhC~\cite{wuSchemeAchievingTopological2015, gorlachFarfieldProbingLeaky2018, parappurathDirectObservationTopological2020, barczykInterplayLeakageRadiation2022}. 
Fig.~\ref{fig:Fig1}c presents the wavevector- and frequency-resolved reflectance for two shrinking factors $\rho=0.99$ and $0.97$, highlighting the substantial enhancement in band visibility achieved through the symmetry breaking mechanism. Here, we recognize pronounced, largely horizontal bands in the photonic dispersion, in a striking departure from the linear Dirac-cone dispersion of the unperturbed lattice. This constitutes the first pivotal result of our study --- the experimental observation of photonic Landau levels in a strained PhC membrane. The clarity with which these states can be resolved is testament to the low loss and large scale of the PhC implementation. The Landau levels are unaffected by the shrinking factor (see Supplemental Material Fig.~\ref{fig:FigED2} for more measurements and simulations with varying $\rho$). They are distributed around the $n=0$ Landau level at $\omega_D \approx 2\pi\cdot 197\,$THz and exhibit Fano lineshapes due to interference with the broad reflection background (see Methods). In the following, if not stated otherwise, the presented reflection measurements are of PhCs with $\rho=0.98$ that have bands which are well visible whilst keeping the symmetry breaking weak. \\
The strain magnitude, as controlled by the parameter $\kappa$, largely affects the energy landscape of the PhC (Fig.~\ref{fig:Fig2}a, see Supplemental Material Fig.~\ref{fig:FigED3} for more measurements and simulations with varying $\kappa$). We see that the bandgap size increases with $\kappa$, and successively more flat states emerge around $\omega_D$. Extracting their center mode frequencies at the $\Gamma$ point from fits to the experimental lineshape (see equation~(\ref{eq:fano}) in Methods), we recognize that the retrieved level separation follows the expected square root scaling that is unique to massless Dirac particles in an increasingly strong external (pseudo-)magnetic field --- see equation~(\ref{eq:landau}). This characteristic energy scaling is further supported by eigenfrequency simulations (Fig.~\ref{fig:Fig2}b). Together, the results of Fig.~\ref{fig:Fig2} underline the origin of the resolved states in the strain-induced PMF.

\begin{figure*}[hbt]
  \centering
  \includegraphics[width=\textwidth]{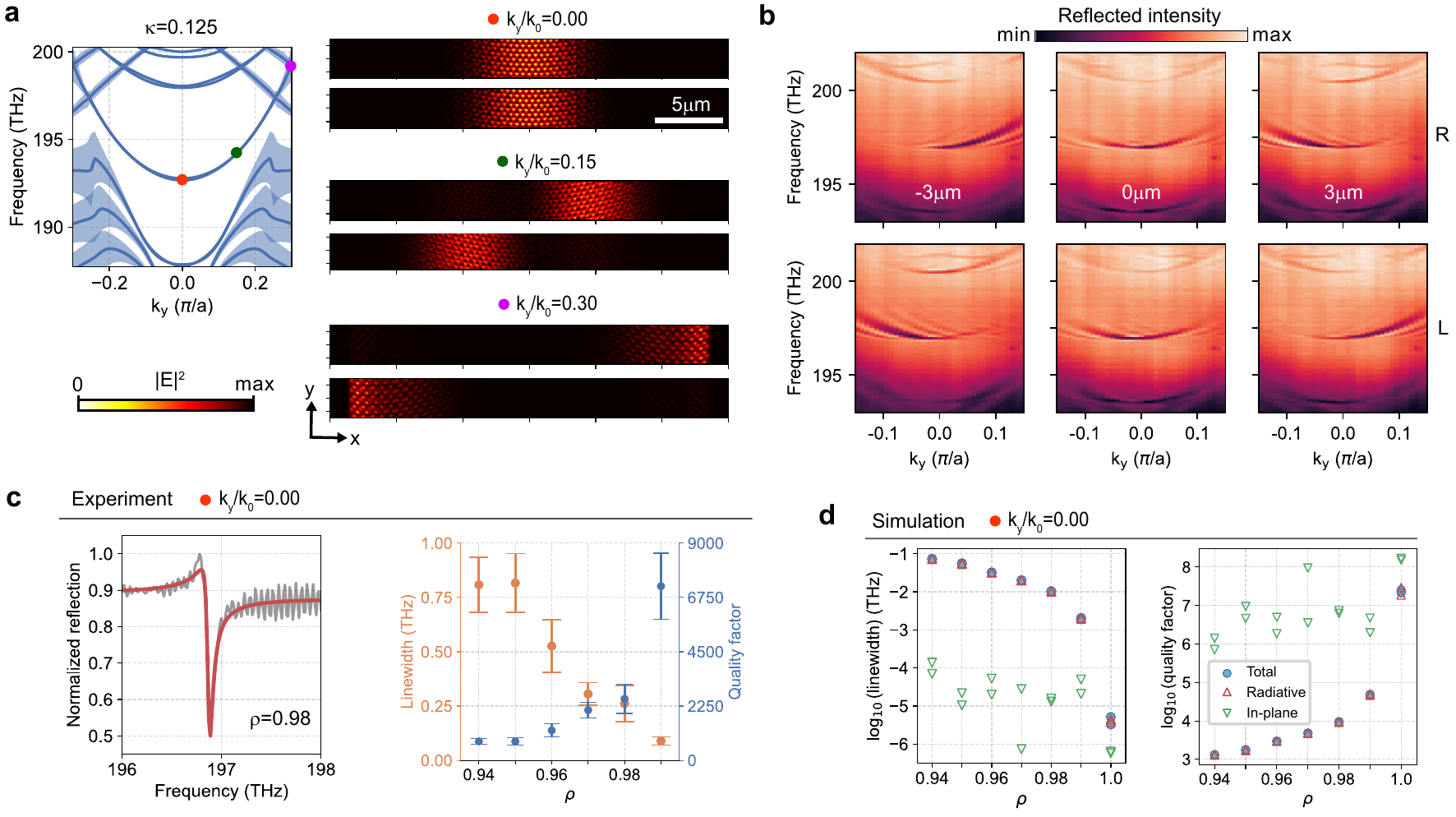}
  \caption{\textbf{Landau level localization, polarization, and losses.} \textbf{a,} Simulated bands of a strained lattice ($\kappa=0.125$, $\rho=0.98$) with the linewidth indicated by blue shading (left), alongside simulated mode profiles of the in-plane electric field intensity for selected values of $k_y$ (right). \textbf{b,} Position- and polarization-dependent excitation of photonic Landau levels, where the displacement in $x$ relative to the lattice center and the right- (R) and left-hand (L) polarization state of the incident beam are indicated. \textbf{c,} Fano-fits to the measured lineshapes (left) yield the experimentally extracted linewidths and quality factors of the $n=0$ Landau level at the $\Gamma$ point as a function of $\rho$ (right). The linewidths are averaged over seven cross cuts of the recorded dispersion along $k_y$ (around $k_y\approx 0$) and given alongside the standard error. \textbf{d,} Numerically retrieved linewidths (left) and quality factors (right), showing the respective contributions of radiative and in-plane losses to the total mode loss.}
  \label{fig:Fig3}
\end{figure*}

\subsection*{Localization and radiation}
Long-range interactions in the PhC lattice render the photonic Landau levels weakly dispersive away from the $\Gamma$ point, increasing in frequency with $|k_y|$ (Fig. \ref{fig:Fig2}a)~\cite{guglielmonLandauLevelsStrained2021}. Coupling beyond next-nearest-neighbors corresponds to higher order contributions ($\mathcal{O}(\kappa^2)$) in equation~(\ref{eq:landau}). It thus represents a feature specific to the photonic platform, differentiating it from the tight-binding graphene analogue. 
The spatial localization of the states also displays a clear dependence on $k_y$. Figure~\ref{fig:Fig3}a shows the simulated fields of the $n=0$ Landau level at three different $k_y$ values. All bands are doubly degenerate, with fields that are even and odd with respect to the mirror symmetry axis of the lattice along $x=0$ --- or an orthogonal pair of superpositions thereof, as we plot in Fig.~\ref{fig:Fig3}a (right). While the states' transverse extent for $k_y=0$ --- where the group velocity vanishes --- spans many unit cells, it appears finite nonetheless. 
For finite wavevector $k_y$, the fields are localized further away from the center, eventually transforming into trivial chiral states that propagate along the armchair edges of the lattice (Fig.~\ref{fig:Fig3}a)~\cite{guglielmonLandauLevelsStrained2021, akhmerovBoundaryConditionsDirac2008, kohmotoZeroModesEdge2007}. 
We see this behavior confirmed in experiment when scanning the excitation focus along $x$, as shown in Fig.~\ref{fig:Fig3}b. In Fig.~\ref{fig:Fig3}b, we plot the measured states for two specific incident polarizations, i.e., right- and left-handed circularly polarized, respectively (see Supplemental Material Fig.~\ref{fig:FigED4} for incident linear polarization).
Interestingly, we recognize that a mode with positive $k_y$ can be excited either with right-circular polarization to the left of the center ($x=-3~\mathrm{\mu m}$) or with left-circular polarization to the right ($x=3~\mathrm{\mu m}$). Thus, these states form a degenerate pair with opposite pseudospin (encoded as far-field helicity) linked to transverse localization. We note that an equivalent pair with opposite handedness exists in the backwards direction, owing to time-reversal symmetry. For excitation near the center $x=0$, the flat part of the band near $k_y=0$ is probed for any polarization.

The transverse localization is also linked to the losses of the states, manifested in their resonance linewidths. We distinguish four possible sources of loss: (i) intrinsic far-field radiation, due to the fact that the states inherently reside within the radiation continuum, (ii) losses associated with the edge of the lattice at large $|x|$, due to either scattering or leakage into slab modes, (iii) losses due to scattering at random disorder, and finally (iv) the radiation losses associated with non-unity shrinking factor $\rho$, which are intentionally introduced to facilitate the free-space measurements.
At the $\Gamma$ point, we  determine the quality factors of the $n=0$ level from linewidth fits (Fig.~\ref{fig:Fig3}c). The measured quality factors reach $Q\approx 7000$ for the weakest symmetry breaking $\rho=0.99$, close to the spectral resolution limit (see Methods). The fact that losses increase strongly with smaller $\rho$ shows that at $k_y=0$ they are limited by the non-intrinsic losses due to non-unity $\rho$. In fact, the linearly decreasing trend of linewidth with $\rho$ shows that losses are negligibly small when extrapolated to $\rho=1$.
Numerical simulations with and without absorbing boundary conditions (see Methods) indeed show that radiative losses dominate (Fig.~\ref{fig:Fig3}d, left), and the intrinsic quality factor ($\rho=1$) is calculated to be on the order of $Q\sim 10^7$ (Fig.~\ref{fig:Fig3}d, right, and Supplemental Material Fig.~\ref{fig:FigED5}a). This is a remarkably high value, given that the photonic Landau levels are inherently residing within the free-space radiation continuum, as can also be seen from numerical simulations of the complex mode frequencies (Fig.~\ref{fig:Fig3}a). So while the strain perturbation strongly alters the real parts of the spectrum, transforming the linear Dirac dispersion into flat Landau levels, it leaves the imaginary parts close to zero.
From the $k_y$-dependent reflection spectra in Fig.~\ref{fig:Fig3}b, we see that resonance linewidths increase away from the $\Gamma$ point. This can be related to the shift in localization away from the lattice center toward the edges, accompanied by increased in-plane losses (ii) that dominate for large wavevectors (see Supplemental Material Fig.~\ref{fig:FigED6} for simulations of the losses versus $k_y$). 

\begin{figure*}[hbt]
  \centering
  \includegraphics[width=\textwidth]{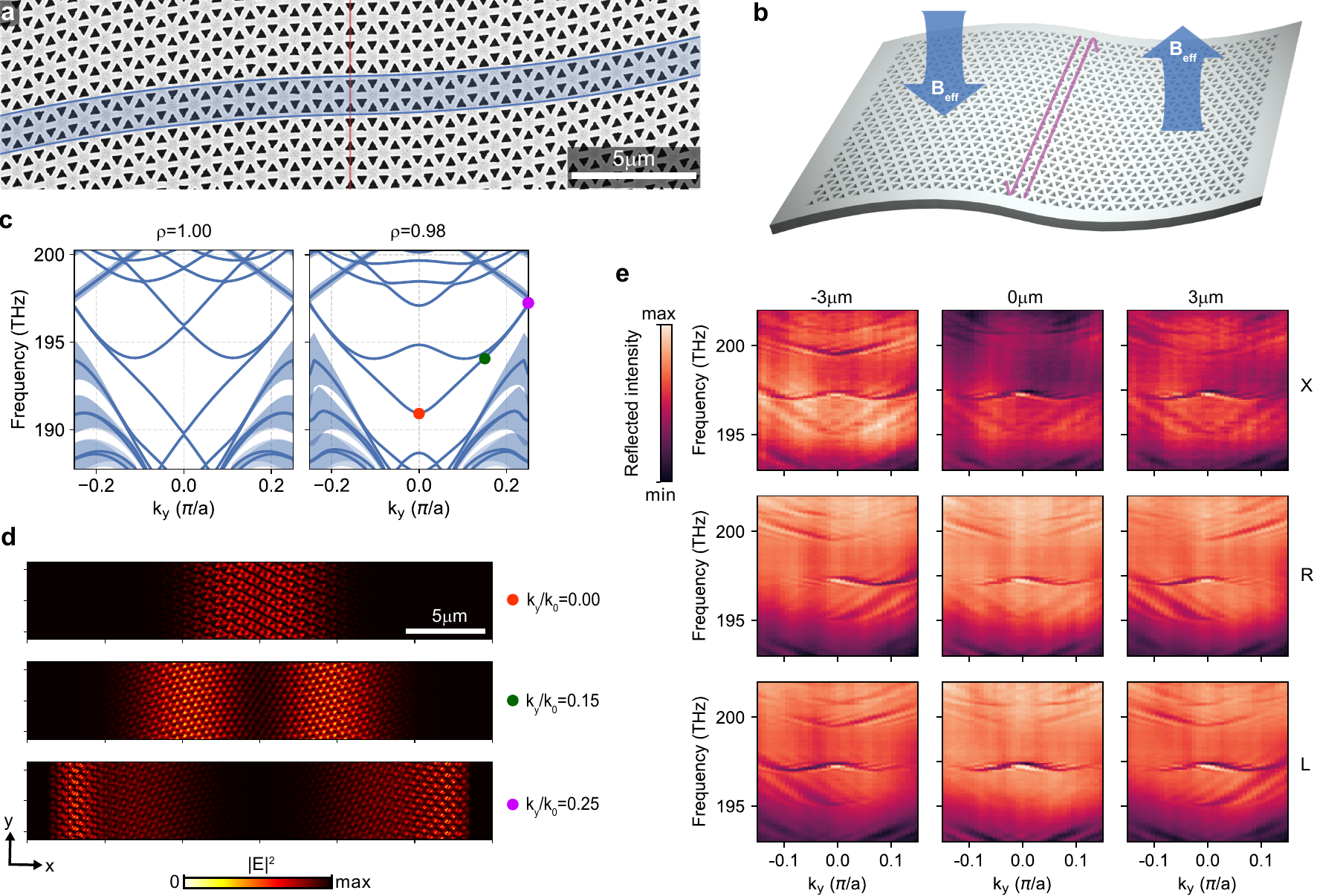}
  \caption{\textbf{Topological edge states through inhomogeneous pseudomagnetic fields.} \textbf{a,} Scanning electron micrograph and \textbf{b,} schematic illustration of a photonic crystal membrane composed of inversely strained domains sharing an armchair interface. The opposing pseudomagnetic fields $\pm B_\text{eff}$ penetrating both domains (blue arrows) lead to the emergence of counterpropagating, spin-polarized topological edge states guided along their mutual boundary (magenta arrows) \textbf{c,} Simulated bands for the pristine ($\rho=1.00$) and shrunken ($\rho=0.98$) case. \textbf{d,} Real-space edge state mode profiles of the in-plane electric field intensity for the values of $k_y$ marked in \textbf{c}. \textbf{e,} Position- and polarization-dependent edge state spectra, where the incident polarization state is denoted as X (linear horizontal), R (right-hand circular), or L (left-hand circular). The strain magnitude is $\kappa=0.125$ for all panels.} 
  \label{fig:Fig4}
\end{figure*}

\subsection*{Topological edge states at domain walls}
A defining advantage of the PhC platform is the ease with which optical potentials can be tailored at will as a function of position. Indeed, different gauge potentials (equation~\ref{eq:aeff}) can be realized through different deformations $\mathbf{u}(\mathbf{x})$. By choosing $\mathbf{u}(\mathbf{x})=(0,a^{-1}(\kappa x)^2\,\text{sgn}\,x)$, i.e., enforcing inversion- rather than mirror-symmetry in the $x=0$ plane of the PhC, we obtain two oppositely strained domains that feature a domain wall between oppositely oriented PMFs at $x=0$ (Fig.~\ref{fig:Fig4}a,b). Numerical simulations predict the emergence of non-degenerate topological edge states at the interface, connecting the former Landau levels of the individual half-domains (Fig.~\ref{fig:Fig4}c, left)~\cite{huangPatterntunableSyntheticGauge2022}. The edge states are localized at the central boundary for $k_y/k_0=0$ but now feature finite group velocity there (Fig.~\ref{fig:Fig4}c, left). Like the Landau levels, they are largely protected from diffraction loss at the PhC terminations, resulting in similarly narrow linewidths and theoretical quality factors up to $10^6$ (Supplemental Material Fig.~\ref{fig:FigED5}b). Away from the $\Gamma$ point, the edge states are increasingly displaced from the central domain wall and slowly restore their degeneracy and the photonic Landau level nature, before eventually also turning into trivial propagating states at the PhC boundaries (Fig.~\ref{fig:Fig4}d), accompanied by enhanced losses and a broadened linewidth (Supplemental Material Fig.~\ref{fig:FigED5}b). The mode localization and associated losses can be broadly tuned by means of changing the PMF magnitude (see Supplemental Material Fig.~\ref{fig:FigED7} for experiments and simulations with varying $\kappa$) or by realizing more complex strain patterns. Figure~\ref{fig:Fig4}e shows Fourier spectroscopy measurements for different excitation positions and polarizations. The symmetry breaking $\rho<1$ we induce to enhance radiative coupling now alters the topological edge state dispersion, introducing avoided crossings at the $\Gamma$ point (Fig.~\ref{fig:Fig4}c, right, see Supplemental Material Fig.~\ref{fig:FigED8} for experiments and simulations with varying $\rho$). Nonetheless, much of the characteristic features of the predicted topological edge states can be recognized in the dispersion bands, as compared to the theory prediction in Fig.~\ref{fig:Fig4}c (right). For right- and left-handed circularly polarized light, we see signatures of spin-orbit coupling in the topological edge states. At a fixed frequency, one can selectively couple into forward or backward propagating modes by changing the helicity (pseudospin) of the incident beam.
However, we note that the same state can be launched with opposite helicity at the other side of the center. Moreover, Fig.~\ref{fig:Fig4}e shows that two edge states can be excited at the same location with equal helicity, despite having opposite group velocity.
Thus, contrary to other implementations of topological edge states in PhCs based on the quantum spin Hall effect \cite{wuSchemeAchievingTopological2015, parappurathDirectObservationTopological2020}, there is no unique correspondence between the states' pseudospin and the far-field helicity of the emitted radiation. 

\section*{Discussion}
We demonstrated the experimental realization of pseudomagnetic fields in PhC membranes via engineered synthetic strain, and employed it to induce photonic Landau levels and topological edge states in the photonic band structure. We studied these states in the far-field by introducing radiative coupling through sub-lattice symmetry breaking. The latter could however be readily removed, leaving remarkably high quality factors despite the states being coupled to the radiation continuum. In fact, these PhC Landau levels share some traits with bound states in the continuum~\cite{hsuBoundStatesContinuum2016}, combining low radiation coupling and spatial delocalization. 
While the predicted extreme ($\sim10^7$) quality factors may not be reached in practice due to random fabrication disorder, the low loss augments the high interest of these states for applications that benefit from strong field enhancement. Together with the slow group velocity and large delocalization, it makes the flat Landau level bands extremely appealing for quantum interfaces, nonlinear nanophotonics, lasers, etc. Moreover, it provides an intriguing testbed to study the physics of Anderson localization.
The reliance on artificial rather than mechanical strain allows for broad engineering of the PhCs' localization and propagation properties, and offers extensive control over dispersion through tailoring of the lattice. For instance, the flatness of the Landau level bands could be readily improved by engineering an effective pseudoelectric field through additional transverse strain along $x$ when exploring applications of light-matter enhancement~\cite{guglielmonLandauLevelsStrained2021}.  
Overall, the demonstrated strain-induced gauge fields provide a new, highly flexible paradigm for the exploration of novel photonic phenomena and the potential development of new photonic devices.

\section*{Methods}

\subsection*{Numerical simulations}
Full-wave finite-element-method simulations in three dimensions were performed using the COMSOL Multiphysics RF Module~\cite{COMSOL52a}. The refractive index of silicon was set to $n = 3.48$, with a slab thickness of $220\,$nm. The unstrained primitive rhombic unit cell consisted of equilateral triangular air holes with side length $s = 0.3125\cdot\sqrt{3}a_0$ and lattice constant $\sqrt{3}a_0 = 800\,$nm. Perfectly matched layers above and to the transverse ($x$) sides of the simulation domain provide us with an estimate for the total loss and associated linewidth of the (quasinormal) eigenmodes, defined as two times the imaginary part of the complex eigenfrequency. We also perform simulations in which the transverse perfectly matched layers are replaced by perfect electric conducting boundaries to eliminate in-plane loss, and thus separately quantify the in-plane and out-of-plane (radiative) contributions to the total loss through comparison. We extract the displayed near-field profiles on a regular grid in a plane located $20\,$nm above the slab.

\subsection*{Device fabrication} 
The PhC slab was fabricated on a silicon-on-insulator platform with a $220\,$nm thick silicon layer on a $3\,\mathrm{\mu m}$ buried oxide layer. First, a positive electron-beam resist of thickness $240\,$nm (AR-P 6200.09) was spin-coated. Then, the PhC design was patterned in the resist using e-beam lithography (Raith Voyager) with $50 \,$kV beam exposure. The e-beam resist was developed in pentyl-acetate/o-Xylene/MIBK:IPA(9:1)/isopropanol, and the chip subsequently underwent reactive-ion etching in HBr and O$_2$. Finally, the buried-oxide layer was removed in an aqueous 4:1 solution of hydrofluoric acid for $19\,$min and the sample was then subjected to critical point drying in order to obtain free-standing PhC membranes.\:\cite{reardonFabricationCharacterizationPhotonic2012} The PhC lattice design features a honeycomb configuration of equilateral triangles (side length $s=0.3125a$) in a hexagonal unit cell with lattice constant $a=827\,$nm. The unit cell shrinking factor was chosen as $\rho=r/r_0=0.98$ unless stated otherwise, whereby $r$ is the distance of the triangular air holes' centroids from the unit cell center, and $r=r_0$ for a perfect honeycomb lattice.

\subsection*{Experimental setup} 
The optical setup is schematically depicted in Supplemental Material Fig.~\ref{fig:FigED1}. To measure the photonic dispersion, we use a $200\,$mW supercontinuum source (SCS, Fianium WhiteLase Micro) that generates light with a broadband spectrum. Its output is filtered by a long-pass filter with a cutoff wavelength of $1150\,$nm and coupled into a single-mode optical fiber. The IR light from the fiber is collimated by an achromatic lens (COL) and passed through a linear polarizer and an achromatic quarter-wave plate, which together define the polarization of the input beam (PO1). A non-polarizing beam splitter cube (BS) steers the input light to an aspheric microscope objective (MO, Olympus LCPLN50XIR, $50\times$, numerical aperture $= 0.65$), which focuses the incident Gaussian beam onto the sample. In order to precisely position the sample in the focal plane, it is attached to a XYZ-movable piezo actuator (MCL Nano-3D200FT, controlled via MCL ND3-USB163), which itself is mounted atop a manual XYZ translation stage for coarse alignment. Reflected light is collected by the same objective and passed through the BS and a second set of linear polarizer and quarter-wave-plate (PO2) to project the BFP radiation onto the desired polarization state. It then passes a Fourier lens (FL) which, together with a tube lens (TL), images the objective's BFP onto the entrance slit of a spectrometer (Acton SpectraPro SP-2300i). Optional custom spatial filters (SFs) are placed in the image plane between the FL and TL to define the sample area from which light is collected and to suppress stray light. The (vertical) entrance slit of the spectrometer is aligned with the optical axis and selects a cross-cut along $k_x = 0$ in the reciprocal plane, confirmed using a test grating sample. With the help of two parabolic mirrors (PMs) for focusing and collection, the spectrometer grating then disperses the broadband IR light orthogonally to the slit, such that the InGaAs IR camera (AVT Goldeye G-008 SWIR) placed at the spectrometer output records images of frequency versus $k_y/k_0$, where $k_0$ is the free-space wave vector. The wave vector resolution is $\delta k_y/k_0 \approx 0.009$, and the typical minimal spectral resolution is $\sim 16\,$GHz.

\subsection*{Extraction of resonance frequencies and quality factors} 
In order to extract the center mode frequencies and quality factors from the reflection spectra, we fit a set of general (Fano) resonance lineshapes of the form
\begin{equation}
\label{eq:fano}
R(\omega) = \left| A_0 + \sum^n_{j=1} A_j e^{i\phi_j} \frac{\gamma_j}{\omega - \omega_{0_j} + i\gamma_j} \right|^2 
\end{equation}
where $A_0$ is a constant background amplitude; and $A_j$, $\phi_j$, and $\omega_{0_j}-i\gamma_j$ are the amplitude, phase, and complex frequency of $n=2$ individual Lorentzians, respectively. One of these Lorentzians models the photonic crystal mode, while the second (broad) Lorentzian accounts for the slowly varying background reflection. Quality factors are defined as $Q_j = \omega_{0_j} / (2\gamma_j)$.

\section*{Data availability}
The main data supporting the findings of this study are available within the article. Additional data and all associated codes for simulations are available from the corresponding authors upon reasonable request.

\section*{Acknowledgements}
We thank Sonakshi Arora and Daniel Muis for fruitful discussions. This work is part of the research programme of the Netherlands Organisation for Scientific Research (NWO). The authors acknowledge support from the European Research Council (ERC) Starting Grant no. 759644-TOPP and Advanced Investigator grant no. 340438-CONSTANS.

\section*{Note}
We would like to note that the group of Mikael Rechtsman has concurrently posted a similar work on the observation of Landau levels in photonic crystals.

\section*{Author Contributions}
R.B. and E.V conceived the project. R.B. fabricated the devices, carried out the measurements, performed data analysis and modeling, and drafted the manuscript. E.V. and L.K. supervised the project. All authors contributed extensively to the interpretation of the results and the writing of the manuscript.

\section*{Competing interests}
The authors declare no competing interests.

\clearpage


\setcounter{section}{0}
\setcounter{equation}{0}
\setcounter{figure}{0}
\setcounter{table}{0}
\setcounter{page}{1}
\renewcommand{\theHfigure}{Supplemental Fig.\,S\thefigure} 
\makeatletter
\renewcommand{\fnum@figure}{\textbf{Fig.\,S\thefigure}}
\let\@keywords\@empty
\makeatother

\title{Supplemental Material: Observation of Landau levels and topological edge states in photonic crystals through pseudomagnetic fields induced by synthetic strain}

\maketitle
\onecolumngrid

\begin{figure}[hbt]
\centering
\includegraphics[width=.6\textwidth]{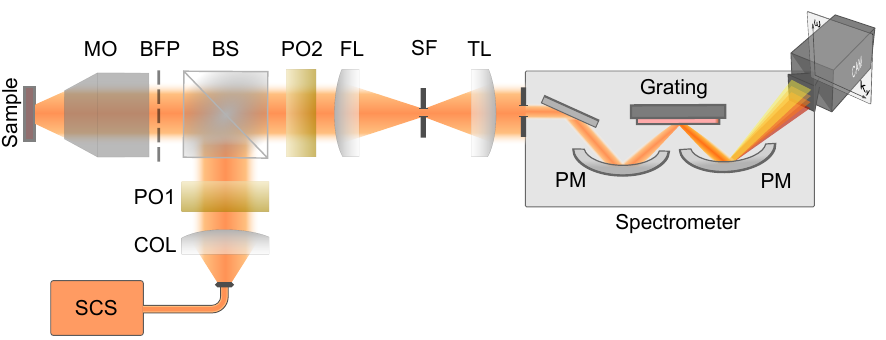}
\caption{\textbf{Far-field Fourier spectroscopy.} Schematic depiction of the experimental far-field Fourier spectropolarimetry setup used for angularly resolved measurement of the photonic crystals' band dispersion. See Methods for details and abbreviations.}
\label{fig:FigED1}
\end{figure}

\begin{figure}[hbt]
\centering
\includegraphics[width=\textwidth]{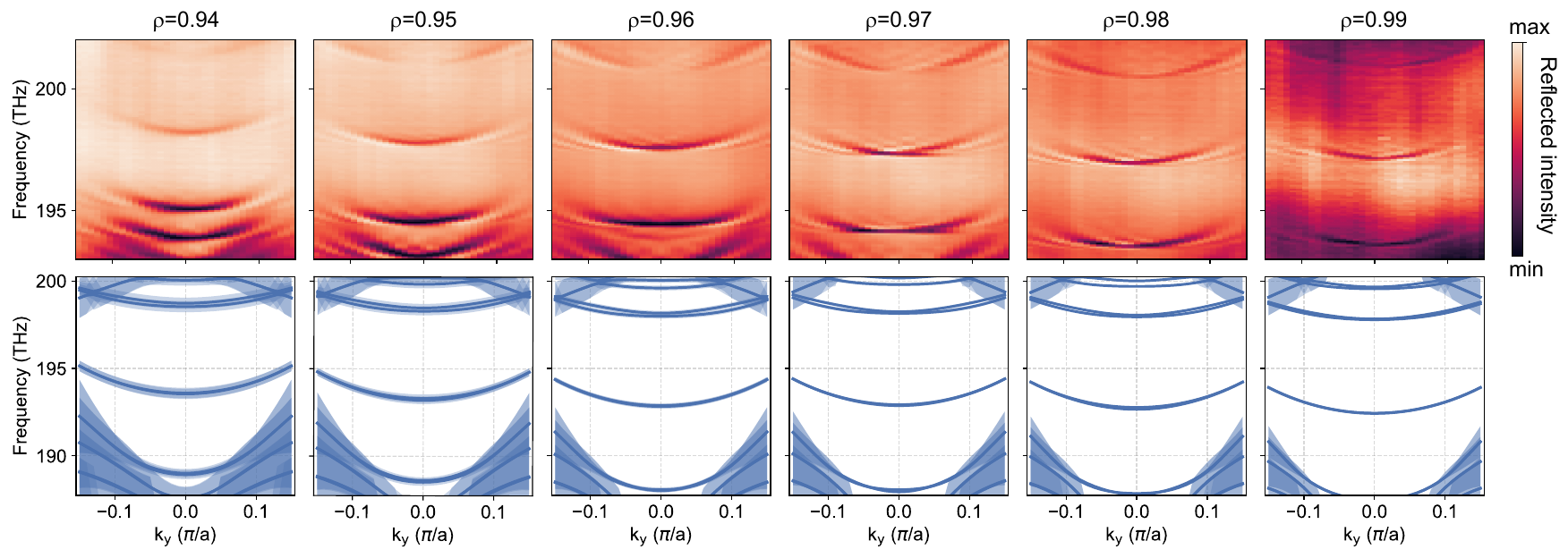}
\caption{\textbf{Symmetry breaking to control radiative coupling.} Measured (top) and simulated (bottom) bands of Landau levels in PhCs with varying unit cell shrinking factor $\rho$, all at $\kappa=0.125$. The linewidth in the simulations is scaled by a factor two for enhanced visibility. The linewidth increases with decreasing $\rho$.}
\label{fig:FigED2}
\end{figure}

\begin{figure}[hbt]
\centering
\includegraphics[width=\textwidth]{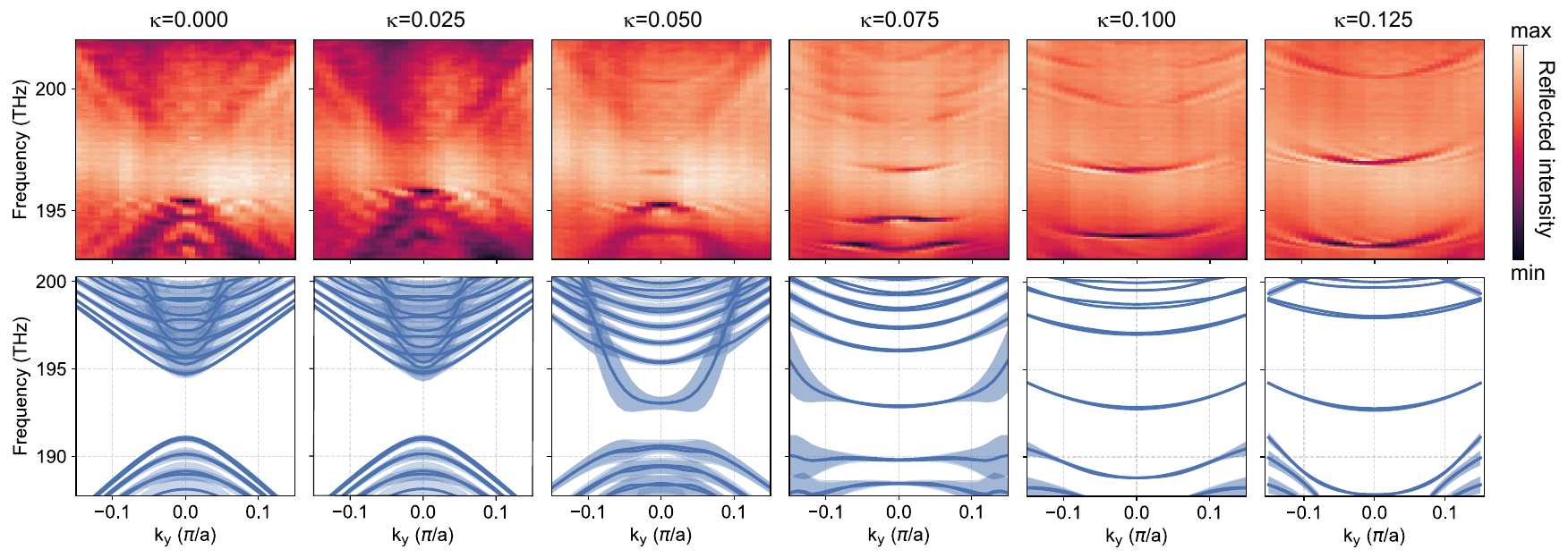}
\caption{\textbf{Tailoring Landau levels via strain.} Measured (top) and simulated (bottom) bands of Landau levels in PhCs with increasing strain magnitude $\kappa$, all at $\rho=0.98$. The gap at small $\kappa$ is due to the sub-lattice symmetry breaking.}
\label{fig:FigED3}
\end{figure}

\begin{figure}[hbt]
\centering
\includegraphics[width=.53\textwidth]{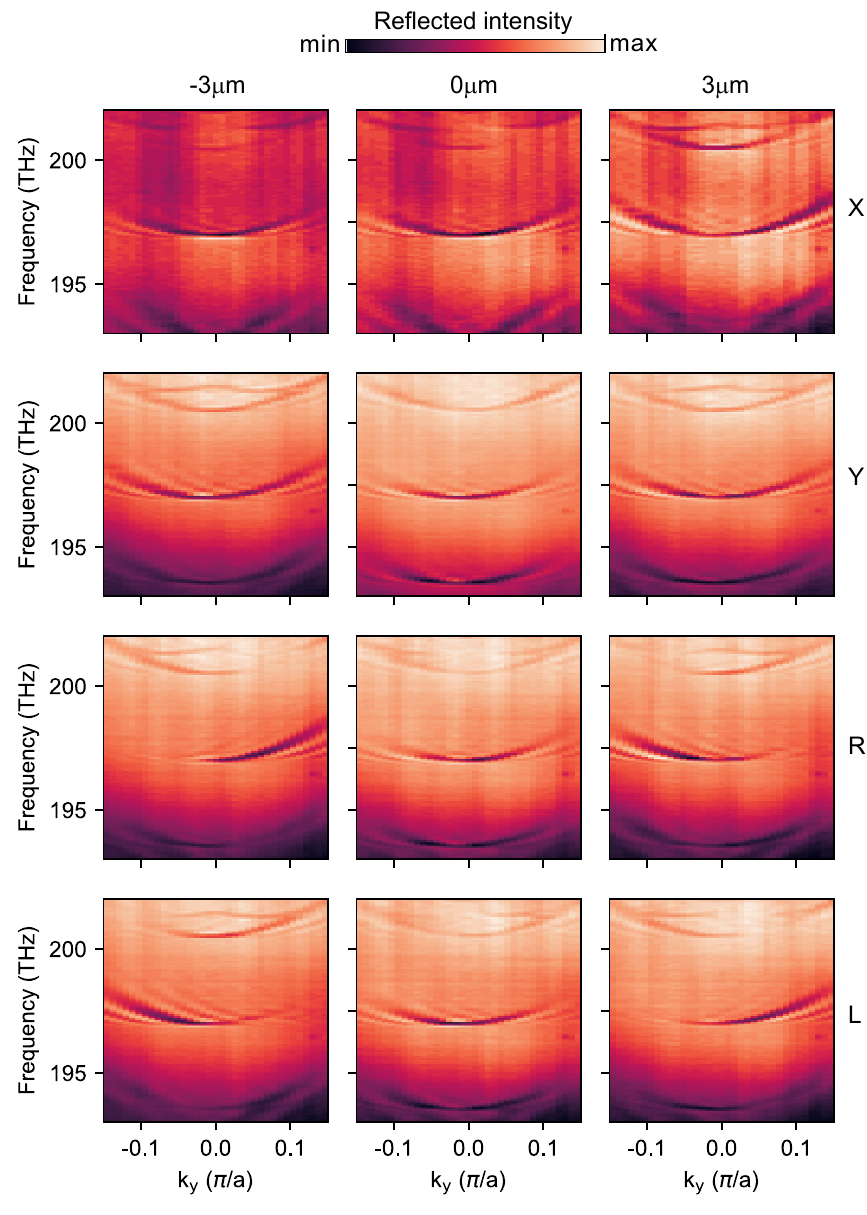}
\caption{\textbf{Landau level localization and polarization.} Position- and polarization-dependent excitation of photonic Landau levels, where the displacement in $x$ relative to the lattice center and the polarization state of the incident beam (linear horizontal (X), linear vertical (Y), right-handed circular (R) and left-handed circular (L)) are indicated ($\rho=0.98$ and $\kappa=0.125$ in these measurements).}
\label{fig:FigED4}
\end{figure}

\begin{figure}[hbt]
\centering
\includegraphics[width=.6\textwidth]{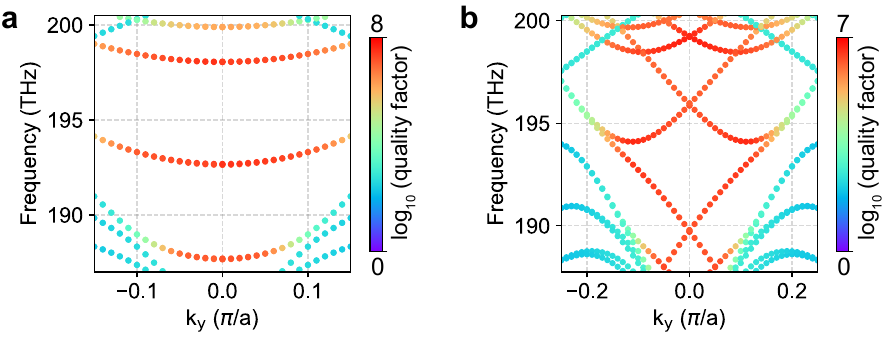}
\caption{\textbf{Quality factors of pristine lattices.} \textbf{a,} Numerically retrieved bands of a pristine strained photonic crystal featuring Landau levels, with color-coded quality factors ($\rho=1.00$, $\kappa=0.125$). \textbf{b,} Same as \textbf{a}, for a photonic crystal featuring topological edge states.}
\label{fig:FigED5}
\end{figure}

\begin{figure}[hbt]
\centering
\includegraphics[width=.6\textwidth]{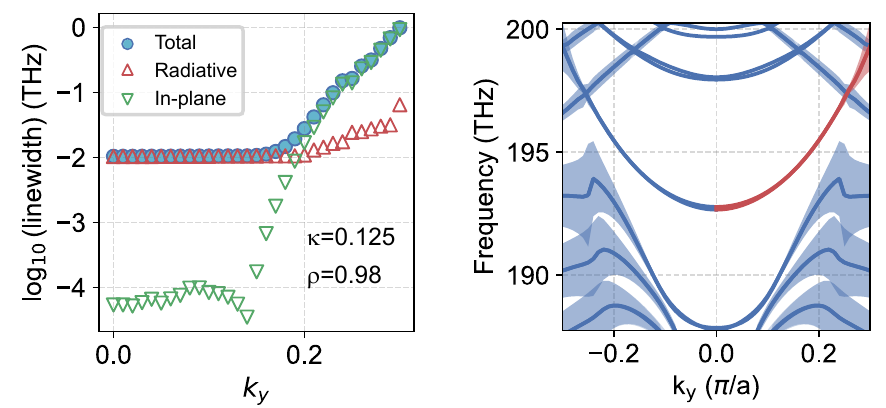}
\caption{\textbf{Landau level losses versus wavevector.} Numerically retrieved contributions of radiative and in-plane losses to the total linewidth of the zeroth Landau level (left), corresponding to the section of the band highlighted in red (right). For large $k_y$, the in-plane losses exceed the radiative losses to the top and bottom of the PhC slab, which reduces the visibility of the bands in the experiment as they become under-coupled to the free-space radiation.}
\label{fig:FigED6}
\end{figure}

\begin{figure}[hbt]
\centering
\includegraphics[width=0.95\textwidth]{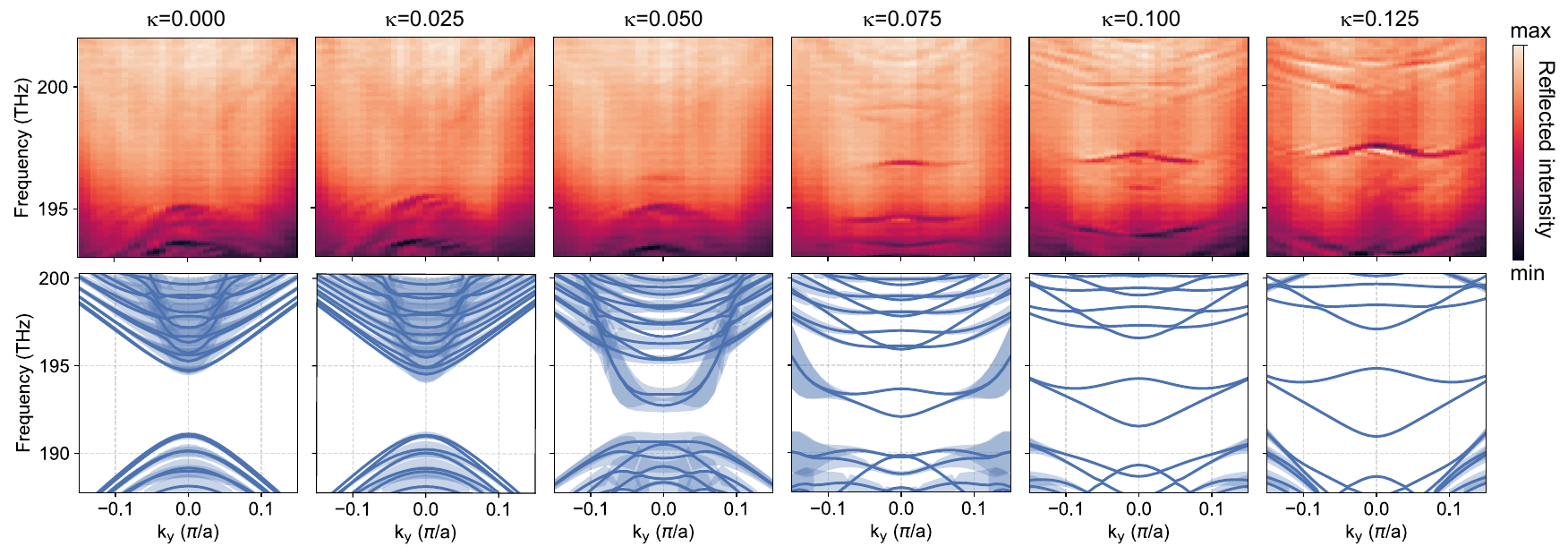}
\caption{\textbf{Tailoring topological edge states via strain.}  Measured (top) and simulated (bottom) bands of topological edge states in PhCs with increasing strain magnitude $\kappa$, all at $\rho=0.98$.}
\label{fig:FigED7}
\end{figure}

\begin{figure}[hbt]
\centering
\includegraphics[width=0.95\textwidth]{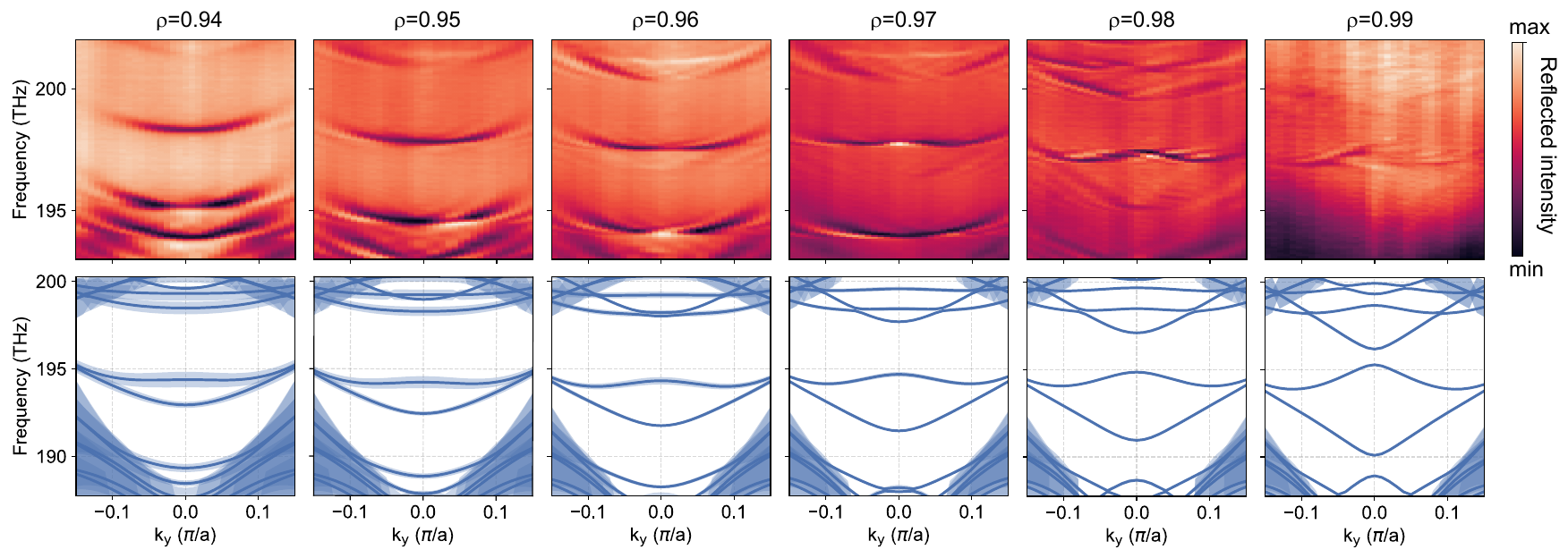}
\caption{\textbf{Sub-lattice symmetry breaking and edge state dispersion.} Measured (top) and simulated (bottom) bands of topological edge states in PhCs with varying unit cell shrinking factor $\rho$, all at $\kappa=0.125$. The linewidth is scaled by a factor two for enhanced visibility. The sub-lattice symmetry breaking leads to avoided crossings around the $\Gamma$-point in the edge state dispersion. }
\label{fig:FigED8}
\end{figure}


\begin{thebibliography}{10}
\expandafter\ifx\csname url\endcsname\relax
  \def\url#1{\texttt{#1}}\fi
\expandafter\ifx\csname urlprefix\endcsname\relax\def\urlprefix{URL }\fi
\providecommand{\bibinfo}[2]{#2}
\providecommand{\eprint}[2][]{\url{#2}}

\bibitem{wangObservationUnidirectionalBackscatteringimmune2009}
\bibinfo{author}{Wang, Z.}, \bibinfo{author}{Chong, Y.},
  \bibinfo{author}{Joannopoulos, J.~D.} \& \bibinfo{author}{Solja{\v c}i{\'c},
  M.}
\newblock \bibinfo{title}{Observation of unidirectional backscattering-immune
  topological electromagnetic states}.
\newblock \emph{\bibinfo{journal}{Nature}} \textbf{\bibinfo{volume}{461}},
  \bibinfo{pages}{772--775} (\bibinfo{year}{2009}).

\bibitem{kaneSizeShapeLow1997}
\bibinfo{author}{Kane, C.~L.} \& \bibinfo{author}{Mele, E.~J.}
\newblock \bibinfo{title}{Size, {{Shape}}, and {{Low Energy Electronic
  Structure}} of {{Carbon Nanotubes}}}.
\newblock \emph{\bibinfo{journal}{Phys. Rev. Lett.}}
  \textbf{\bibinfo{volume}{78}}, \bibinfo{pages}{1932--1935}
  (\bibinfo{year}{1997}).

\bibitem{guineaEnergyGapsZerofield2010}
\bibinfo{author}{Guinea, F.}, \bibinfo{author}{Katsnelson, M.~I.} \&
  \bibinfo{author}{Geim, A.~K.}
\newblock \bibinfo{title}{Energy gaps and a zero-field quantum {{Hall}} effect
  in graphene by strain engineering}.
\newblock \emph{\bibinfo{journal}{Nat. Phys.}} \textbf{\bibinfo{volume}{6}},
  \bibinfo{pages}{30--33} (\bibinfo{year}{2010}).

\bibitem{levyStrainInducedPseudoMagnetic2010}
\bibinfo{author}{Levy, N.} \emph{et~al.}
\newblock \bibinfo{title}{Strain-{{Induced Pseudo}}\textendash{{Magnetic Fields
  Greater Than}} 300 {{Tesla}} in {{Graphene Nanobubbles}}}.
\newblock \emph{\bibinfo{journal}{Science}} \textbf{\bibinfo{volume}{329}},
  \bibinfo{pages}{544--547} (\bibinfo{year}{2010}).

\bibitem{gomesDesignerDiracFermions2012}
\bibinfo{author}{Gomes, K.~K.}, \bibinfo{author}{Mar, W.}, \bibinfo{author}{Ko,
  W.}, \bibinfo{author}{Guinea, F.} \& \bibinfo{author}{Manoharan, H.~C.}
\newblock \bibinfo{title}{Designer {{Dirac}} fermions and topological phases in
  molecular graphene}.
\newblock \emph{\bibinfo{journal}{Nature}} \textbf{\bibinfo{volume}{483}},
  \bibinfo{pages}{306--310} (\bibinfo{year}{2012}).

\bibitem{rechtsmanStraininducedPseudomagneticField2013}
\bibinfo{author}{Rechtsman, M.~C.} \emph{et~al.}
\newblock \bibinfo{title}{Strain-induced pseudomagnetic field and photonic
  {{Landau}} levels in dielectric structures}.
\newblock \emph{\bibinfo{journal}{Nat. Photon.}} \textbf{\bibinfo{volume}{7}},
  \bibinfo{pages}{153--158} (\bibinfo{year}{2013}).

\bibitem{songDispersionlessCouplingOptical2022}
\bibinfo{author}{Song, W.} \emph{et~al.}
\newblock \bibinfo{title}{Dispersionless {{Coupling}} among {{Optical
  Waveguides}} by {{Artificial Gauge Field}}}.
\newblock \emph{\bibinfo{journal}{Phys. Rev. Lett.}}
  \textbf{\bibinfo{volume}{129}}, \bibinfo{pages}{053901}
  (\bibinfo{year}{2022}).

\bibitem{bellecObservationSupersymmetricPseudoLandau2020}
\bibinfo{author}{Bellec, M.}, \bibinfo{author}{Poli, C.},
  \bibinfo{author}{Kuhl, U.}, \bibinfo{author}{Mortessagne, F.} \&
  \bibinfo{author}{Schomerus, H.}
\newblock \bibinfo{title}{Observation of supersymmetric pseudo-{{Landau}}
  levels in strained microwave graphene}.
\newblock \emph{\bibinfo{journal}{Light Sci. Appl.}}
  \textbf{\bibinfo{volume}{9}}, \bibinfo{pages}{146} (\bibinfo{year}{2020}).

\bibitem{jiaExperimentalRealizationChiral2022}
\bibinfo{author}{Jia, H.} \emph{et~al.}
\newblock \bibinfo{title}{Experimental realization of chiral {{Landau}} levels
  in two-dimensional {{Dirac}} cone systems with inhomogeneous effective mass}.
\newblock \emph{\bibinfo{journal}{\emph{Preprint at
  https://arxiv.org/abs/2209.10745}}}  (\bibinfo{year}{2022}).

\bibitem{jamadiDirectObservationPhotonic2020}
\bibinfo{author}{Jamadi, O.} \emph{et~al.}
\newblock \bibinfo{title}{Direct observation of photonic {{Landau}} levels and
  helical edge states in strained honeycomb lattices}.
\newblock \emph{\bibinfo{journal}{Light Sci. Appl.}}
  \textbf{\bibinfo{volume}{9}}, \bibinfo{pages}{144} (\bibinfo{year}{2020}).

\bibitem{lledoPolaritonCondensationVortex2022}
\bibinfo{author}{Lled{\'o}, C.}, \bibinfo{author}{Carusotto, I.} \&
  \bibinfo{author}{Szymanska, M.}
\newblock \bibinfo{title}{Polariton condensation into vortex states in the
  synthetic magnetic field of a strained honeycomb lattice}.
\newblock \emph{\bibinfo{journal}{SciPost Phys.}}
  \textbf{\bibinfo{volume}{12}}, \bibinfo{pages}{068} (\bibinfo{year}{2022}).

\bibitem{zhengLandauLevelsVan2021}
\bibinfo{author}{Zheng, S.} \emph{et~al.}
\newblock \bibinfo{title}{Landau {{Levels}} and van der {{Waals Interfaces}} of
  {{Acoustics}} in {{Moir}}\'e {{Phononic Lattices}}}.
\newblock \emph{\bibinfo{journal}{\emph{Preprint at
  https://arxiv.org/abs/2209.10745}}}  (\bibinfo{year}{2021}).

\bibitem{wenAcousticLandauQuantization2019}
\bibinfo{author}{Wen, X.} \emph{et~al.}
\newblock \bibinfo{title}{Acoustic {{Landau}} quantization and
  quantum-{{Hall-like}} edge states}.
\newblock \emph{\bibinfo{journal}{Nat. Phys.}} \textbf{\bibinfo{volume}{15}},
  \bibinfo{pages}{352--356} (\bibinfo{year}{2019}).

\bibitem{yangStrainInducedGaugeField2017}
\bibinfo{author}{Yang, Z.}, \bibinfo{author}{Gao, F.}, \bibinfo{author}{Yang,
  Y.} \& \bibinfo{author}{Zhang, B.}
\newblock \bibinfo{title}{Strain-{{Induced Gauge Field}} and {{Landau Levels}}
  in {{Acoustic Structures}}}.
\newblock \emph{\bibinfo{journal}{Phys. Rev. Lett.}}
  \textbf{\bibinfo{volume}{118}}, \bibinfo{pages}{194301}
  (\bibinfo{year}{2017}).

\bibitem{abbaszadehSonicLandauLevels2017}
\bibinfo{author}{Abbaszadeh, H.}, \bibinfo{author}{Souslov, A.},
  \bibinfo{author}{Paulose, J.}, \bibinfo{author}{Schomerus, H.} \&
  \bibinfo{author}{Vitelli, V.}
\newblock \bibinfo{title}{Sonic {{Landau Levels}} and {{Synthetic Gauge
  Fields}} in {{Mechanical Metamaterials}}}.
\newblock \emph{\bibinfo{journal}{Phys. Rev. Lett.}}
  \textbf{\bibinfo{volume}{119}}, \bibinfo{pages}{195502}
  (\bibinfo{year}{2017}).

\bibitem{periAxialfieldinducedChiralChannels2019}
\bibinfo{author}{Peri, V.}, \bibinfo{author}{{Serra-Garcia}, M.},
  \bibinfo{author}{Ilan, R.} \& \bibinfo{author}{Huber, S.~D.}
\newblock \bibinfo{title}{Axial-field-induced chiral channels in an acoustic
  {{Weyl}} system}.
\newblock \emph{\bibinfo{journal}{Nat. Phys.}} \textbf{\bibinfo{volume}{15}},
  \bibinfo{pages}{357--361} (\bibinfo{year}{2019}).

\bibitem{guglielmonLandauLevelsStrained2021}
\bibinfo{author}{Guglielmon, J.}, \bibinfo{author}{Rechtsman, M.~C.} \&
  \bibinfo{author}{Weinstein, M.~I.}
\newblock \bibinfo{title}{Landau levels in strained two-dimensional photonic
  crystals}.
\newblock \emph{\bibinfo{journal}{Phys. Rev. A}}
  \textbf{\bibinfo{volume}{103}}, \bibinfo{pages}{013505}
  (\bibinfo{year}{2021}).

\bibitem{salernoHowDirectlyObserve2015}
\bibinfo{author}{Salerno, G.}, \bibinfo{author}{Ozawa, T.},
  \bibinfo{author}{Price, H.~M.} \& \bibinfo{author}{Carusotto, I.}
\newblock \bibinfo{title}{How to directly observe {{Landau}} levels in
  driven-dissipative strained honeycomb lattices}.
\newblock \emph{\bibinfo{journal}{2D Mater.}} \textbf{\bibinfo{volume}{2}},
  \bibinfo{pages}{034015} (\bibinfo{year}{2015}).

\bibitem{salernoPropagatingEdgeStates2017}
\bibinfo{author}{Salerno, G.}, \bibinfo{author}{Ozawa, T.},
  \bibinfo{author}{Price, H.~M.} \& \bibinfo{author}{Carusotto, I.}
\newblock \bibinfo{title}{Propagating edge states in strained honeycomb
  lattices}.
\newblock \emph{\bibinfo{journal}{Phys. Rev. B}} \textbf{\bibinfo{volume}{95}},
  \bibinfo{pages}{245418} (\bibinfo{year}{2017}).

\bibitem{borregaardQuantumNetworksDeterministic2019}
\bibinfo{author}{Borregaard, J.}, \bibinfo{author}{S{\o}rensen, A.~S.} \&
  \bibinfo{author}{Lodahl, P.}
\newblock \bibinfo{title}{Quantum {{Networks}} with {{Deterministic
  Spin}}\textendash{{Photon Interfaces}}}.
\newblock \emph{\bibinfo{journal}{Adv. Quantum Technol.}}
  \textbf{\bibinfo{volume}{2}}, \bibinfo{pages}{1800091}
  (\bibinfo{year}{2019}).

\bibitem{kraussWhyWeNeed2008}
\bibinfo{author}{Krauss, T.~F.}
\newblock \bibinfo{title}{Why do we need slow light?}
\newblock \emph{\bibinfo{journal}{Nat. Photon.}} \textbf{\bibinfo{volume}{2}},
  \bibinfo{pages}{448--450} (\bibinfo{year}{2008}).

\bibitem{smirnovaNonlinearTopologicalPhotonics2020}
\bibinfo{author}{Smirnova, D.}, \bibinfo{author}{Leykam, D.},
  \bibinfo{author}{Chong, Y.} \& \bibinfo{author}{Kivshar, Y.}
\newblock \bibinfo{title}{Nonlinear topological photonics}.
\newblock \emph{\bibinfo{journal}{Appl. Phys. Rev.}}
  \textbf{\bibinfo{volume}{7}}, \bibinfo{pages}{021306} (\bibinfo{year}{2020}).

\bibitem{schomerusParityAnomalyLandauLevel2013}
\bibinfo{author}{Schomerus, H.} \& \bibinfo{author}{Halpern, N.~Y.}
\newblock \bibinfo{title}{Parity {{Anomaly}} and {{Landau-Level Lasing}} in
  {{Strained Photonic Honeycomb Lattices}}}.
\newblock \emph{\bibinfo{journal}{Phys. Rev. Lett.}}
  \textbf{\bibinfo{volume}{110}}, \bibinfo{pages}{013903}
  (\bibinfo{year}{2013}).

\bibitem{yangPhotonicFlatbandResonances2023}
\bibinfo{author}{Yang, Y.} \emph{et~al.}
\newblock \bibinfo{title}{Photonic flatband resonances for free-electron
  radiation}.
\newblock \emph{\bibinfo{journal}{Nature}} \textbf{\bibinfo{volume}{613}},
  \bibinfo{pages}{42--47} (\bibinfo{year}{2023}).

\bibitem{gorlachFarfieldProbingLeaky2018}
\bibinfo{author}{Gorlach, M.~A.} \emph{et~al.}
\newblock \bibinfo{title}{Far-field probing of leaky topological states in
  all-dielectric metasurfaces}.
\newblock \emph{\bibinfo{journal}{Nat. Commun.}} \textbf{\bibinfo{volume}{9}},
  \bibinfo{pages}{909} (\bibinfo{year}{2018}).

\bibitem{parappurathDirectObservationTopological2020}
\bibinfo{author}{Parappurath, N.}, \bibinfo{author}{Alpeggiani, F.},
  \bibinfo{author}{Kuipers, L.} \& \bibinfo{author}{Verhagen, E.}
\newblock \bibinfo{title}{Direct observation of topological edge states in
  silicon photonic crystals: {{Spin}}, dispersion, and chiral routing}.
\newblock \emph{\bibinfo{journal}{Sci. Adv.}} \textbf{\bibinfo{volume}{6}},
  \bibinfo{pages}{eaaw4137} (\bibinfo{year}{2020}).

\bibitem{barczykInterplayLeakageRadiation2022}
\bibinfo{author}{Barczyk, R.} \emph{et~al.}
\newblock \bibinfo{title}{Interplay of {{Leakage Radiation}} and {{Protection}}
  in {{Topological Photonic Crystal Cavities}}}.
\newblock \emph{\bibinfo{journal}{Laser Photonics Rev.}}
  \textbf{\bibinfo{volume}{2022}}, \bibinfo{pages}{2200071}
  (\bibinfo{year}{2022}).

\bibitem{huangPatterntunableSyntheticGauge2022}
\bibinfo{author}{Huang, Z.-T.} \emph{et~al.}
\newblock \bibinfo{title}{Pattern-tunable synthetic gauge fields in topological
  photonic graphene}.
\newblock \emph{\bibinfo{journal}{Nanophotonics}}  (\bibinfo{year}{2022}).

\bibitem{wuSchemeAchievingTopological2015}
\bibinfo{author}{Wu, L.-H.} \& \bibinfo{author}{Hu, X.}
\newblock \bibinfo{title}{Scheme for {{Achieving}} a {{Topological Photonic
  Crystal}} by {{Using Dielectric Material}}}.
\newblock \emph{\bibinfo{journal}{Phys. Rev. Lett.}}
  \textbf{\bibinfo{volume}{114}}, \bibinfo{pages}{223901}
  (\bibinfo{year}{2015}).

\bibitem{kiriushechkinaSpindependentPropertiesOptical2023}
\bibinfo{author}{Kiriushechkina, S.} \emph{et~al.}
\newblock \bibinfo{title}{Spin-dependent properties of optical modes guided by
  adiabatic trapping potentials in photonic {{Dirac}} metasurfaces}.
\newblock \emph{\bibinfo{journal}{Nat. Nanotechnol.}} \bibinfo{pages}{1--7}
  (\bibinfo{year}{2023}).

\bibitem{renZeroenergyEdgeStates2023}
\bibinfo{author}{Ren, B.} \emph{et~al.}
\newblock \bibinfo{title}{Zero-energy edge states and solitons in strained
  photonic graphene}.
\newblock \emph{\bibinfo{journal}{Phys. Rev. A}}
  \textbf{\bibinfo{volume}{107}}, \bibinfo{pages}{043504}
  (\bibinfo{year}{2023}).

\bibitem{barikTwodimensionallyConfinedTopological2016}
\bibinfo{author}{Barik, S.}, \bibinfo{author}{Miyake, H.},
  \bibinfo{author}{DeGottardi, W.}, \bibinfo{author}{Waks, E.} \&
  \bibinfo{author}{Hafezi, M.}
\newblock \bibinfo{title}{Two-dimensionally confined topological edge states in
  photonic crystals}.
\newblock \emph{\bibinfo{journal}{New J. Phys.}} \textbf{\bibinfo{volume}{18}},
  \bibinfo{pages}{113013} (\bibinfo{year}{2016}).

\bibitem{reardonFabricationCharacterizationPhotonic2012}
\bibinfo{author}{Reardon, C.~P.}, \bibinfo{author}{Rey, I.~H.},
  \bibinfo{author}{Welna, K.}, \bibinfo{author}{O'Faolain, L.} \&
  \bibinfo{author}{Krauss, T.~F.}
\newblock \bibinfo{title}{Fabrication {{And Characterization Of Photonic
  Crystal Slow Light Waveguides And Cavities}}}.
\newblock \emph{\bibinfo{journal}{J. Vis. Exp.}}  (\bibinfo{year}{2012}).

\bibitem{akhmerovBoundaryConditionsDirac2008}
\bibinfo{author}{Akhmerov, A.~R.} \& \bibinfo{author}{Beenakker, C. W.~J.}
\newblock \bibinfo{title}{Boundary conditions for {{Dirac}} fermions on a
  terminated honeycomb lattice}.
\newblock \emph{\bibinfo{journal}{Phys. Rev. B}} \textbf{\bibinfo{volume}{77}},
  \bibinfo{pages}{085423} (\bibinfo{year}{2008}).

\bibitem{kohmotoZeroModesEdge2007}
\bibinfo{author}{Kohmoto, M.} \& \bibinfo{author}{Hasegawa, Y.}
\newblock \bibinfo{title}{Zero modes and edge states of the honeycomb lattice}.
\newblock \emph{\bibinfo{journal}{Phys. Rev. B}} \textbf{\bibinfo{volume}{76}},
  \bibinfo{pages}{205402} (\bibinfo{year}{2007}).

\bibitem{hsuBoundStatesContinuum2016}
\bibinfo{author}{Hsu, C.~W.}, \bibinfo{author}{Zhen, B.},
  \bibinfo{author}{Stone, A.~D.}, \bibinfo{author}{Joannopoulos, J.~D.} \&
  \bibinfo{author}{Solja{\v c}i{\'c}, M.}
\newblock \bibinfo{title}{Bound states in the continuum}.
\newblock \emph{\bibinfo{journal}{Nat. Rev. Mater.}}
  \textbf{\bibinfo{volume}{1}}, \bibinfo{pages}{1--13} (\bibinfo{year}{2016}).

\bibitem{COMSOL52a}
\bibinfo{author}{{COMSOL Multiphysics® v. 5.2.}}
\newblock \emph{\bibinfo{title}{{www.comsol.com COMSOL AB}}}
  (\bibinfo{publisher}{Stockholm}, \bibinfo{address}{Sweden},
  \bibinfo{year}{2015}).

\end{thebibliography}
\end{document}